\documentclass[copyright,creativecommons]{eptcs}
\providecommand{\event}{QPL 2011}

\providecommand{\titlerunning}{On the Automation of Encoding Processes in the Quantum IO Monad}
\providecommand{\authorrunning}{James Barratt}

\usepackage[utf8]{inputenc} 

\usepackage{graphicx} 

\usepackage{booktabs} 
\usepackage{array} 
\usepackage{paralist} 
\usepackage{verbatim} 
\usepackage{subfig} 

\makeatletter
\@ifundefined{lhs2tex.lhs2tex.sty.read}%
  {\@namedef{lhs2tex.lhs2tex.sty.read}{}%
   \newcommand\SkipToFmtEnd{}%
   \newcommand\EndFmtInput{}%
   \long\def\SkipToFmtEnd#1\EndFmtInput{}%
  }\SkipToFmtEnd

\newcommand\ReadOnlyOnce[1]{\@ifundefined{#1}{\@namedef{#1}{}}\SkipToFmtEnd}
\usepackage{amstext}
\usepackage{amssymb}
\usepackage{stmaryrd}
\DeclareFontFamily{OT1}{cmtex}{}
\DeclareFontShape{OT1}{cmtex}{m}{n}
  {<5><6><7><8>cmtex8
   <9>cmtex9
   <10><10.95><12><14.4><17.28><20.74><24.88>cmtex10}{}
\DeclareFontShape{OT1}{cmtex}{m}{it}
  {<-> ssub * cmtt/m/it}{}

\DeclareFontShape{OT1}{cmtt}{bx}{n}
  {<5><6><7><8>cmtt8
   <9>cmbtt9
   <10><10.95><12><14.4><17.28><20.74><24.88>cmbtt10}{}
\DeclareFontShape{OT1}{cmtex}{bx}{n}
  {<-> ssub * cmtt/bx/n}{}

\newcommand{\Conid}[1]{\mathit{#1}}
\newcommand{\Varid}[1]{\mathit{#1}}
\newcommand{\anonymous}{\kern0.06em \vbox{\hrule\@width.5em}}

\usepackage{polytable}

\@ifundefined{mathindent}%
  {\newdimen\mathindent\mathindent\leftmargini}%
  {}%

\def\resethooks{%
  \global\let\SaveRestoreHook\empty
  \global\let\ColumnHook\empty}
\newcommand*{\savecolumns}[1][default]%
  {\g@addto@macro\SaveRestoreHook{\savecolumns[#1]}}
\newcommand*{\restorecolumns}[1][default]%
  {\g@addto@macro\SaveRestoreHook{\restorecolumns[#1]}}
\newcommand*{\aligncolumn}[2]%
  {\g@addto@macro\ColumnHook{\column{#1}{#2}}}

\resethooks

\newcommand{\onelinecommentchars}{\quad-{}- }
\newcommand{\commentbeginchars}{\enskip\{-}
\newcommand{\commentendchars}{-\}\enskip}

\newcommand{\visiblecomments}{%
  \let\onelinecomment=\onelinecommentchars
  \let\commentbegin=\commentbeginchars
  \let\commentend=\commentendchars}

\newcommand{\invisiblecomments}{%
  \let\onelinecomment=\empty
  \let\commentbegin=\empty
  \let\commentend=\empty}

\visiblecomments

\newlength{\blanklineskip}
\newcommand{\hsindent}[1]{\quad}
\let\hspre\empty
\let\hspost\empty

\EndFmtInput
\makeatother
\title{On the Automation of Encoding Processes in the Quantum IO Monad}
\author{James Barratt
\institute{Rule Financial Ltd}
\email{james.barratt@rulefinancial.co.uk}
}

\def\titlerunning{Encoding Processes in the QIO Monad}

\begin{document}
\maketitle

\begin{abstract}
It is now clear that the use of resilient encoding schemes will be required for any quantum computing device to be realised. However, quantum programmers of the future will not wish to be tied up in the particulars of such encoding schemes. Quantum programming languages and libraries are already being developed, one of which is the Quantum IO Monad. QIO, as it is often abbreviated to, provides an interface to define and simulate quantum computations via a library of functions written in Haskell, a purely functional programming language. A solution is presented that takes an arbitrary QIO program and returns an equivalent program incorporating some specified quantum error correction techniques.
\end{abstract}

\section{Introduction}

Although theoretically possible, it is not yet clear whether building a sizeable quantum computer is feasible. The main obstacle is to find ways of dealing with decoherence and other quantum noise. When a quantum system interacts with its environment quantum information is leaked out and it begins to act in a probabilistic manner. In effect, parts of the system are being measured by the environment. Extraneous operations may also appear randomly during the computation, which corrupts the state and ultimately the result. Although research is focused on building quantum computers that are less likely to interact with their environment, it is impossible to completely isolate a system and therefore decoherence and quantum noise are inevitable. However, a number of techniques have been developed by the quantum software community that promise to reduce their effects on the system even further.

Quantum error correction encapsulates software related techniques for reducing the impact of decoherence and other quantum noise. Some of the techniques that have so far been developed are inspired by coding theory and techniques already used within classical error correction. The basic principle is to encode information in such a way that the existence of errors can be detected and the nature of those errors identified. It is then possible to apply recovery operators before the information is decoded into its original form. However, checking for errors in a quantum computer is more problematic and there are new, entirely non classical, errors to contend with.

\section{The Quantum IO Monad}

The Quantum IO Monad is a library of functions written in Haskell that provides an interface to define and simulate quantum computations\cite{Altenkirch}. It is used within the current work to demonstrate the use of encoding schemes. In functional programming computation is considered to be the application of functions as opposed to the manipulation of some global state. The following is a simple example of a QIO program.
\smallskip
\begingroup\par\noindent\advance\leftskip\mathindent\(
\begin{pboxed}\SaveRestoreHook
\column{B}{@{}>{\hspre}l<{\hspost}@{}}%
\column{3}{@{}>{\hspre}l<{\hspost}@{}}%
\column{E}{@{}>{\hspre}l<{\hspost}@{}}%
\>[B]{}\Varid{example}\mathbin{::}\Conid{QIO}\;\Conid{Bool}{}\<[E]%
\\
\>[B]{}\Varid{example}\mathrel{=}\mathbf{do}{}\<[E]%
\\
\>[B]{}\hsindent{3}{}\<[3]%
\>[3]{}\Varid{q1}\leftarrow \Varid{mkQbit}\;\Conid{False}{}\<[E]%
\\
\>[B]{}\hsindent{3}{}\<[3]%
\>[3]{}\Varid{applyU}\mathbin{\$}\Varid{unot}\;\Varid{q1}{}\<[E]%
\\
\>[B]{}\hsindent{3}{}\<[3]%
\>[3]{}\Varid{b}\leftarrow \Varid{measQbit}\;\Varid{q1}{}\<[E]%
\\
\>[B]{}\hsindent{3}{}\<[3]%
\>[3]{}\Varid{return}\;\Varid{b}{}\<[E]%
\ColumnHook
\end{pboxed}
\)\par\noindent\endgroup\resethooks

This function is written using \textbf{do} notation, which resembles a more imperative style, but is in fact functional underneath. Such seemingly impure computations may be defined thanks to the use of monads that are built upon some polymorphic type. The function \textit{mkQbit} creates a new qubit initialised in the state given by the boolean argument, with False and True representing the base states $ | 0 \rangle $ and $ | 1 \rangle $ respectively. Unitary operations are then applied using \textit{applyU}, the quantum NOT gate, or the Pauli X gate, in this case. The \textit{unot} operation is defined as a \textit{Rotation}, as are all single qubit operations. Other unitaries defined in QIO include \textit{swap}, \textit{cond,} for controlled operations, and \textit{ulet}, which is used to declare and encapsulate the use of ancillary qubits. Finally, qubits are measured using \textit{measQbit} and a boolean value is returned randomly depending on the superposition of the qubit at the time. The qubit itself is also collapsed into the base state corresponding to this boolean value.

\section{Implementing Quantum Error Correction in QIO}

Although it may be some time before real quantum computers are realised, work is already being carried out by computer scientists to develop good quantum programming languages. The development of the Quantum IO Monad is an example of such work. Although reliable error correction techniques will be crucial, in order to allow programmers to concentrate on the matter at hand some form of automated conversion into equivalent but error resilient programs would be extremely advantageous. The following describes an attempt to do just this within the Quantum IO Monad.

\subsection{Introducing an Encoded Qubit Type}

The first step towards being able to convert a QIO program into an equivalent that incorporates quantum error correction is to declare a class called \textit{EnQubit}, instances of which will represent encoded qubits. By defining such a class the main program can be written in a generic fashion without the need to understand how encoded qubits are represented\cite{Barratt}. As long as the instances of this class provide definitions of the functions declared here there will be no need to alter the main program when switching schemes, since the appropriate functions will be called at run time.

Each \textit{EnQbit} is thought to have a "parent" qubit, the one from which the encoded state is produced. The function \textit{getQbit} returns this particular qubit and is required by \textit{measEnQbit}, defined below.
\smallskip
\begingroup\par\noindent\advance\leftskip\mathindent\(
\begin{pboxed}\SaveRestoreHook
\column{B}{@{}>{\hspre}l<{\hspost}@{}}%
\column{5}{@{}>{\hspre}l<{\hspost}@{}}%
\column{18}{@{}>{\hspre}l<{\hspost}@{}}%
\column{E}{@{}>{\hspre}l<{\hspost}@{}}%
\>[B]{}\mathbf{class}\;\Conid{EnQbit}\;\Varid{a}\;\mathbf{where}{}\<[E]%
\\
\>[B]{}\hsindent{5}{}\<[5]%
\>[5]{}\Varid{mkEnQbit}\mathbin{::}\Conid{Bool}\to \Conid{QIO}\;\Varid{a}{}\<[E]%
\\
\>[B]{}\hsindent{5}{}\<[5]%
\>[5]{}\Varid{getQbit}\mathbin{::}\Varid{a}\to \Conid{Qbit}{}\<[E]%
\\
\>[B]{}\hsindent{5}{}\<[5]%
\>[5]{}\Varid{measEnQbit}\mathbin{::}\Varid{a}\to \Conid{QIO}\;\Conid{Bool}{}\<[E]%
\\
\>[B]{}\hsindent{5}{}\<[5]%
\>[5]{}\Varid{measEnQbit}\;\Varid{eq}\mathrel{=}\mathbf{do}{}\<[E]%
\\
\>[5]{}\hsindent{13}{}\<[18]%
\>[18]{}\Varid{applyU}\mathbin{\$}\Varid{decode}\;\Varid{eq}{}\<[E]%
\\
\>[5]{}\hsindent{13}{}\<[18]%
\>[18]{}\Varid{b}\leftarrow \Varid{measQbit}\;(\Varid{getQbit}\;\Varid{eq}){}\<[E]%
\\
\>[5]{}\hsindent{13}{}\<[18]%
\>[18]{}\Varid{applyU}\mathbin{\$}\Varid{encode}\;\Varid{eq}{}\<[E]%
\\
\>[5]{}\hsindent{13}{}\<[18]%
\>[18]{}\Varid{return}\;\Varid{b}{}\<[E]%
\ColumnHook
\end{pboxed}
\)\par\noindent\endgroup\resethooks

The class also provides flexibility in the error correcting codes that may be employed. By providing different \textit{encode} and \textit{correct} functions a different encoding scheme may be used with the same representation. The function \textit{decode} is in fact defined here as the inverse of \textit{encode}, since it is unitary.
\smallskip
\begingroup\par\noindent\advance\leftskip\mathindent\(
\begin{pboxed}\SaveRestoreHook
\column{B}{@{}>{\hspre}l<{\hspost}@{}}%
\column{5}{@{}>{\hspre}l<{\hspost}@{}}%
\column{E}{@{}>{\hspre}l<{\hspost}@{}}%
\>[B]{}\hsindent{5}{}\<[5]%
\>[5]{}\Varid{encode}\mathbin{::}\Varid{a}\to \Conid{U}{}\<[E]%
\\
\>[B]{}\hsindent{5}{}\<[5]%
\>[5]{}\Varid{decode}\mathbin{::}\Varid{a}\to \Conid{U}{}\<[E]%
\\
\>[B]{}\hsindent{5}{}\<[5]%
\>[5]{}\Varid{decode}\;\Varid{eq}\mathrel{=}\Varid{urev}\mathbin{\$}\Varid{encode}\;\Varid{eq}{}\<[E]%
\\
\>[B]{}\hsindent{5}{}\<[5]%
\>[5]{}\Varid{correct}\mathbin{::}\Varid{a}\to \Conid{U}{}\<[E]%
\ColumnHook
\end{pboxed}
\)\par\noindent\endgroup\resethooks

The following functions define \textit{EnQbit} versions of the standard unitary operators of QIO. These are implemented to simply decode the given \textit{EnQbit} and call the appropriate standard unitary, passing the "parent" qubit. Before returning, the qubits are encoded again. This approach only provides protection in between the application of operations, while qubits are perhaps being stored or transmitted, as suggested by Shor\cite{Shor}. Being able to perform operations on the actual encoded qubits themselves would make decoding unnecessary until the end of the computation and thus provide greater protection. This is possible for certain operations in a bitwise fashion, those with the property of transversality, but it does depend on the code being used\cite{Zurek}. So although these functions are defined here as a standard, they may be overridden by instances of the class, if such encoded operations are available.
\smallskip
\begingroup\par\noindent\advance\leftskip\mathindent\(
\begin{pboxed}\SaveRestoreHook
\column{B}{@{}>{\hspre}l<{\hspost}@{}}%
\column{5}{@{}>{\hspre}l<{\hspost}@{}}%
\column{9}{@{}>{\hspre}l<{\hspost}@{}}%
\column{E}{@{}>{\hspre}l<{\hspost}@{}}%
\>[B]{}\hsindent{5}{}\<[5]%
\>[5]{}\Varid{rotEnQbit}\mathbin{::}\Varid{a}\to \Conid{Rotation}\to \Conid{U}{}\<[E]%
\\
\>[B]{}\hsindent{5}{}\<[5]%
\>[5]{}\Varid{swapEnQbit}\mathbin{::}\Varid{a}\to \Varid{a}\to \Conid{U}{}\<[E]%
\\
\>[B]{}\hsindent{5}{}\<[5]%
\>[5]{}\Varid{condEnQbit}\mathbin{::}\Varid{a}\to (\Conid{Bool}\to \Conid{U})\to \Conid{U}{}\<[E]%
\\
\>[B]{}\hsindent{5}{}\<[5]%
\>[5]{}\Varid{uletEnQbit}\mathbin{::}\Conid{Bool}\to (\Conid{Qbit}\to \Varid{a}\to \Conid{U})\to \Conid{U}{}\<[E]%
\ColumnHook
\end{pboxed}
\)\par\noindent\endgroup\resethooks

The \textit{uletEnQbit} function is not defined here but simply declared as this depends entirely on the number of qubits being used in the encoding, to determine how many ancillas to create.

In terms of encoded qubit representations, the following type could be used for the three qubit bit flip code, with \textit{EnQbit} functions defined as appropriate.
\smallskip
\begingroup\par\noindent\advance\leftskip\mathindent\(
\begin{pboxed}\SaveRestoreHook
\column{B}{@{}>{\hspre}l<{\hspost}@{}}%
\column{5}{@{}>{\hspre}l<{\hspost}@{}}%
\column{E}{@{}>{\hspre}l<{\hspost}@{}}%
\>[B]{}\mathbf{newtype}\;\Conid{EQ3}\mathrel{=}\Conid{EQ3}\;(\Conid{Qbit},\Conid{Qbit},\Conid{Qbit}){}\<[E]%
\\[\blanklineskip]%
\>[B]{}\mathbf{instance}\;\Conid{EnQbit}\;\Conid{EQ3}\;\mathbf{where}{}\<[E]%
\\
\>[B]{}\hsindent{5}{}\<[5]%
\>[5]{}\Varid{mkEnQbit}\mathrel{=}\Varid{mkEQ3}{}\<[E]%
\\
\>[B]{}\hsindent{5}{}\<[5]%
\>[5]{}\Varid{getQbit}\mathrel{=}\Varid{fstEQ3}{}\<[E]%
\\
\>[B]{}\hsindent{5}{}\<[5]%
\>[5]{}\Varid{encode}\mathrel{=}\Varid{encode3}{}\<[E]%
\\
\>[B]{}\hsindent{5}{}\<[5]%
\>[5]{}\Varid{correct}\mathrel{=}\Varid{correct3}{}\<[E]%
\\
\>[B]{}\hsindent{5}{}\<[5]%
\>[5]{}\Varid{uletEnQbit}\mathrel{=}\Varid{uletEQ3}{}\<[E]%
\ColumnHook
\end{pboxed}
\)\par\noindent\endgroup\resethooks

However, this representation could also be used for the three qubit phase flip code by simply altering the implementations of \textit{encode} and \textit{correct}. An instance has also been implemented for Steane's code\cite{Steane} using a 7-tuple of qubits and appropriate encoding and correction procedures.

\subsection{Converting a QIO program}

In order to manipulate an existing QIO program, the function \textit{ecQIO'} pattern matches on it and inserts new function calls depending on the construct found, exploiting the monadic structure of the \textit{QIO} type. It is defined recursively with the base case being \textit{QReturn}, signifying the end of the computation.
\smallskip
\begingroup\par\noindent\advance\leftskip\mathindent\(
\begin{pboxed}\SaveRestoreHook
\column{B}{@{}>{\hspre}l<{\hspost}@{}}%
\column{E}{@{}>{\hspre}l<{\hspost}@{}}%
\>[B]{}\Varid{ecQIO'}\mathbin{::}\Conid{EnQbit}\;\Varid{a}\Rightarrow \Conid{QIO}\;\Varid{t}\to [\mskip1.5mu \Varid{a}\mskip1.5mu]\to \Conid{QIO}\;\Varid{t}{}\<[E]%
\\
\>[B]{}\Varid{ecQIO'}\;(\Conid{QReturn}\;\Varid{a})\;\anonymous \mathrel{=}\Conid{QReturn}\;\Varid{a}{}\<[E]%
\ColumnHook
\end{pboxed}
\)\par\noindent\endgroup\resethooks

When a qubit is being created, measured or manipulated the appropriate \textit{EnQbit} function is substituted in to replace the original. As such we obtain a program that consists of extra operations and qubits, together encapsulating the logical state of the single qubit referred to. Whatsmore, a list is passed on in the recursive call and threaded through the whole computation as a register of the encoded qubits. In the case of \textit{MkQbit} the newly created \textit{EnQbit} is appended to the list before passing it on. Here \textit{MkQbit} is one of the constructors of the \textit{QIO} type, which the convenience function \textit{mkQbit} seen earlier relies upon.
\smallskip
\begingroup\par\noindent\advance\leftskip\mathindent\(
\begin{pboxed}\SaveRestoreHook
\column{B}{@{}>{\hspre}l<{\hspost}@{}}%
\column{3}{@{}>{\hspre}l<{\hspost}@{}}%
\column{E}{@{}>{\hspre}l<{\hspost}@{}}%
\>[B]{}\Varid{ecQIO'}\;(\Conid{MkQbit}\;\Varid{b}\;\Varid{f})\;\Varid{eqs}\mathrel{=}\mathbf{do}{}\<[E]%
\\
\>[B]{}\hsindent{3}{}\<[3]%
\>[3]{}\Varid{eq}\leftarrow \Varid{mkEnQbit}\;\Varid{b}{}\<[E]%
\\
\>[B]{}\hsindent{3}{}\<[3]%
\>[3]{}\Varid{ecQIO'}\;(\Varid{f}\;(\Varid{getQbit}\;\Varid{eq}))\;(\Varid{eq}\mathbin{:}\Varid{eqs}){}\<[E]%
\ColumnHook
\end{pboxed}
\)\par\noindent\endgroup\resethooks

The interesting case in \textit{ecQIO'} is the \textit{ApplyU} constructor, where it is necessary to pattern match further on the \textit{U} data type representing unitary operations. This takes place in the function \textit{extendU}, where we plan to extend the operation's influence to include the rest of the qubits in the encoded qubit. 
\smallskip
\begingroup\par\noindent\advance\leftskip\mathindent\(
\begin{pboxed}\SaveRestoreHook
\column{B}{@{}>{\hspre}l<{\hspost}@{}}%
\column{3}{@{}>{\hspre}l<{\hspost}@{}}%
\column{E}{@{}>{\hspre}l<{\hspost}@{}}%
\>[B]{}\Varid{ecQIO'}\;(\Conid{ApplyU}\;\Varid{u}\;\Varid{f})\;\Varid{eqs}\mathrel{=}\mathbf{do}{}\<[E]%
\\
\>[B]{}\hsindent{3}{}\<[3]%
\>[3]{}\Varid{applyU}\mathbin{\$}\Varid{extendU}\;\Varid{u}\;\Varid{eqs}{}\<[E]%
\\
\>[B]{}\hsindent{3}{}\<[3]%
\>[3]{}\Varid{ecQIO'}\;\Varid{f}\;\Varid{eqs}{}\<[E]%
\ColumnHook
\end{pboxed}
\)\par\noindent\endgroup\resethooks

By pattern matching on the \textit{U} data type being applied, \textit{extendU} calls the appropriate function implementing the \textit{EnQbit} version of the unitary. The case of a rotation is given here.
\smallskip
\begingroup\par\noindent\advance\leftskip\mathindent\(
\begin{pboxed}\SaveRestoreHook
\column{B}{@{}>{\hspre}l<{\hspost}@{}}%
\column{5}{@{}>{\hspre}l<{\hspost}@{}}%
\column{9}{@{}>{\hspre}l<{\hspost}@{}}%
\column{E}{@{}>{\hspre}l<{\hspost}@{}}%
\>[B]{}\Varid{extendU}\mathbin{::}\Conid{EnQbit}\;\Varid{a}\Rightarrow \Conid{U}\to [\mskip1.5mu \Varid{a}\mskip1.5mu]\to \Conid{U}{}\<[E]%
\\
\>[B]{}\Varid{extendU}\;(\Conid{Rot}\;\Varid{q}\;\Varid{r}\;\Varid{u})\;\Varid{eqs}\mathrel{=}{}\<[E]%
\\
\>[B]{}\hsindent{5}{}\<[5]%
\>[5]{}\mathbf{let}\;\Varid{eq}\mathrel{=}(\Varid{fromJust}\;(\Varid{getEnQbit}\;\Varid{q}\;\Varid{eqs}))\;\mathbf{in}{}\<[E]%
\\
\>[B]{}\hsindent{5}{}\<[5]%
\>[5]{}\Varid{rotEnQbit}\;\Varid{eq}\;\Varid{r}\mathbin{`\Varid{mappend}`}{}\<[E]%
\\
\>[B]{}\hsindent{5}{}\<[5]%
\>[5]{}\Varid{correctAll}\;\Varid{eqs}\mathbin{`\Varid{mappend}`}{}\<[E]%
\\
\>[B]{}\hsindent{5}{}\<[5]%
\>[5]{}\Varid{extendU}\;\Varid{u}\;\Varid{eqs}{}\<[E]%
\ColumnHook
\end{pboxed}
\)\par\noindent\endgroup\resethooks

Since a standard QIO program performs operations on individual qubits, a mechanism for identifying their associated \textit{EnQbit} is needed when faced with measurement and operator constructs. A function \textit{getEnQbit} performs this task by making a comparison between the qubit given and the "parent" qubit of each \textit{EnQbit} in our list.

The function \textit{correctAll} is called upon, and the returned unitary appended to the end of the rotation, to perform the error correction procedure on all encoded qubits created so far, which should take place periodically in order to prevent a build up of errors that is unmanageable for the code. If this were to happen the decoding procedure would yield an incorrect value.

\section{Conclusions}

In terms of providing error correction to QIO programs, the components that are responsible for converting a program appear to fulfill their task of inserting new function calls at appropriate points. Flexibility is also provided on the implementation of codes used, as the same converting functions can work with any coding scheme.

Two encoding schemes have been implemented, the first being the three qubit bit flip code, with Steane's code as the second. Unfortunately, it appears that programs encoded with Steane's code are not evaluated efficiently, preventing any real observation of adding Steane's code to even the simplest of programs. As more qubits and/or unitary operations are added the performance degrades very quickly. This is not such an issue for the three qubit bit flip code, which is evaluated fairly quickly even for larger programs using error correction.

The applicability of this kind of encoding automation has been demonstrated however, and if these evaluations were to be taking place on a real quantum computer then these problems would not be present.

\section*{Acknowledgements}

I would like to thank my MSc dissertation supervisor, Thorsten Altenkirch, for his guidance throughout the course of this work, and also Alexander Green, for his assistance.

I would also like to thank my Mother and Father for their unwavering support, and for simply putting up with me.

Lastly, I would like to dedicate this paper to the memory of my brother, Bill Barratt (1972-2010).

\bibliographystyle{eptcs}
\bibliography{qpl}

\begin{thebibliography}{1}
\providecommand{\bibitemdeclare}[2]{}
\providecommand{\urlprefix}{Available at }
\providecommand{\url}[1]{\texttt{#1}}
\providecommand{\href}[2]{\texttt{#2}}
\providecommand{\urlalt}[2]{\href{#1}{#2}}
\providecommand{\doi}[1]{doi:\urlalt{http://dx.doi.org/#1}{#1}}
\providecommand{\bibinfo}[2]{#2}

\bibitemdeclare{incollection}{Altenkirch}
\bibitem{Altenkirch}
\bibinfo{author}{Thorsten Altenkirch} \& \bibinfo{author}{Alexander Green}
  (\bibinfo{year}{2010}): \emph{\bibinfo{title}{The {Quantum} {IO} {Monad}}}.
\newblock In \bibinfo{editor}{Simon Gay} \& \bibinfo{editor}{Ian McKie},
  editors: {\sl \bibinfo{booktitle}{Semantic Techniques in Quantum
  Computation}}, \bibinfo{publisher}{Cambridge University Press}, pp.
  \bibinfo{pages}{173--205}.

\bibitemdeclare{mastersthesis}{Barratt}
\bibitem{Barratt}
\bibinfo{author}{James Barratt} (\bibinfo{year}{2010}):
  \emph{\bibinfo{title}{Simulating Decoherence and Implementing Quantum Error
  Correction in the Quantum IO Monad}}.
\newblock Master's thesis, \bibinfo{school}{University of Nottingham}.

\bibitemdeclare{article}{Shor}
\bibitem{Shor}
\bibinfo{author}{Peter~W. Shor} (\bibinfo{year}{1995}):
  \emph{\bibinfo{title}{Scheme for reducing decoherence in quantum computer
  memory}}.
\newblock {\sl \bibinfo{journal}{Phys. Rev. A}} \bibinfo{volume}{52}, pp.
  \bibinfo{pages}{R2493--R2496}, \doi{10.1103/PhysRevA.52.R2493}.

\bibitemdeclare{article}{Steane}
\bibitem{Steane}
\bibinfo{author}{A.~M. Steane} (\bibinfo{year}{1996}):
  \emph{\bibinfo{title}{Error Correcting Codes in Quantum Theory}}.
\newblock {\sl \bibinfo{journal}{Phys. Rev. Lett.}} \bibinfo{volume}{77}, pp.
  \bibinfo{pages}{793--797}, \doi{10.1103/PhysRevLett.77.793}.

\bibitemdeclare{article}{Zurek}
\bibitem{Zurek}
\bibinfo{author}{Wojciech~Hubert Zurek} \& \bibinfo{author}{Raymond Laflamme}
  (\bibinfo{year}{1996}): \emph{\bibinfo{title}{Quantum Logical Operations on
  Encoded Qubits}}.
\newblock {\sl \bibinfo{journal}{Phys. Rev. Lett.}} \bibinfo{volume}{77}, pp.
  \bibinfo{pages}{4683--4686}, \doi{10.1103/PhysRevLett.77.4683}.

\end{thebibliography}

\end{document}